\documentclass[10pt,conference]{IEEEtran}
\usepackage{amsmath,amssymb,amsfonts}
\usepackage{algorithmic}
\usepackage{balance}
\usepackage{etoolbox}
\usepackage{graphicx}
\usepackage[breaklinks]{hyperref}
\usepackage{textcomp}
\usepackage{xcolor}
\usepackage[T1]{fontenc} 
\usepackage{threeparttable}

\begin{document}

\newcommand\RQone{\textit{What existing analysis techniques are suitable for identifying areas of improvement in Agile software engineering project teams' communication and collaboration?}}
\newcommand\RQtwo{\textit{What features in project management tools can be used to address areas of improvement in team communication and collaboration?}}
\newcommand\RQthree{\textit{How does providing these features affect student and team performance?}}

\newcommand{\edit}[1]{{\color{blue}{{}#1}}}

\title{Improving Software Team Communication Through Social Interventions in Project Management Tools}

\author{\IEEEauthorblockN{
April Clarke}
\IEEEauthorblockA{\textit{Computer Science and Software Engineering} \\
\textit{University of Canterbury}\\
Christchurch, New Zealand \\
april.clarke@pg.canterbury.ac.nz}}

\maketitle

\begin{abstract}
Productive software engineering teams require effective communication and balanced contributions between team members. However, teams are often ineffective at these skills, which is detrimental to project success. Project-based university courses are an opportunity for students to practise these skills, but we have yet to establish how we can guide students towards improving their communication and coordination. We aim to develop project management tool features, informed by social network analysis, that nudge students in software engineering group projects towards beneficial behaviours. To do this, we will first evaluate the suitability of social network analysis techniques for identifying areas of improvement in teams' communication. Then, we will develop features in a project management tool that aid students in identifying and addressing these areas of improvement, and evaluate them in the context of a software engineering group project.
\end{abstract}

\section{Introduction}
Communication and balanced contributions between team members are widely recognised as key components of effective teamwork, and are crucial for the success of software engineering teams~\cite{curtisFieldStudySoftware1988,hoeglTeamworkQualitySuccess2001,mensahImpactTeamworkQuality2024}. However, teams are often ineffective at these skills~\cite{krautCoordinationSoftwareDevelopment1995}, leading to reduced productivity and effectiveness~\cite{r.noelExploringCollaborativeWriting2018,strodeTeamworkEffectivenessModel2022a}, which is detrimental to project success. In particular, teams identify a lack of knowledge sharing within the team and communication breakdowns as sources of issues, especially in times of stress~\cite{salasHowCanYou1997}.


An anti-pattern that contributes to ineffective knowledge sharing is the presence of ``communication focal points'', which occur when knowledgeable team members who are skilled communicators act as communication brokers in their teams~\cite{omalleyAnalysisSocialNetworks2008}. While these members are recognised by their teammates as being exceptional contributors to the team, dependence on communication brokers causes stress due to their investment in the team's success~\cite{curtisFieldStudySoftware1988}, and reduces performance at the team level~\cite{sparroweSocialNetworksPerformance2001}. In well-performing teams, members tend to know each others' expertise~\cite{r.noelExploringCollaborativeWriting2018}, so they are able to coordinate knowledge sharing without relying on communication brokers. Teams with these coordination skills not only show better performance, but also higher team member satisfaction~\cite{mensahImpactTeamworkQuality2024,seersTeammemberExchangeQuality1989}, indicating that improving communication and coordination benefits both individuals and teams.

To identify communication behaviours that can negatively affect team productivity, such as reliance on communication brokers, we can use social network analysis (SNA)~\cite{otteSocialNetworkAnalysis2002}. Social networks represent relationships between entities, where each node in the network is an entity, and each edge between two nodes represents a relationship. Analysing the characteristics of these networks is a means to identify communication patterns that cause productivity differences between teams. Social networks are often based on written or verbal communication, but in the software engineering context, we can also use data like bug reports, files, or projects developers contribute to~\cite{zhangUnderstandingStudentTeachers2022}. 


While there is extensive research on the characteristics of effective teams and the importance of communication~\cite{mathieuTeamEffectiveness199720072008}, there is little work regarding \textit{how} we can improve social networks based on SNA. We propose that SNA can identify areas of improvement in teams' communication, and these areas of improvement can inform interventions at a student level that encourage behaviours that strengthen team communication and coordination. Such behaviours may include members who need to coordinate working together so the stress on communication brokers is reduced, or creating opportunities for knowledge sharing and improving quiet team members' self-confidence by contributing to key features and pair programming with each of their teammates~\cite{hughesRemotePairProgramming2020, mcdowellPairProgrammingImproves2006}.

We aim to assist students in targeting their effort towards improving communication and coordination skills through automated interventions in project management tools. To do this, we will first identify methods of nudging them towards positive behaviours. We will implement our proposed design in a project management tool, and iterate to improve it over two instances of an Agile software engineering group project course. The intended outcome of this research is better intra-team communication, which will improve team performance and member satisfaction.

To guide this, we will answer the following questions:
\begin{enumerate}
    \item RQ1: \RQone
    \item RQ2: \RQtwo
    \item RQ3: \RQthree
\end{enumerate}

The immediate benefit of the proposed work will be improving education in software engineering group projects through targeting communication-related learning outcomes.



\section{Related Work}

Betweenness centrality represents how frequently a node is found in the shortest path between any other two nodes. O'Malley and Marsden~\cite{omalleyAnalysisSocialNetworks2008} found that nodes with higher betweenness centrality acted as communication brokers in networks, and tended to control relationships between other nodes. Once this anti-pattern has been identified in a team, a corrective intervention could encourage task allocations that spread knowledge across more teammates.

Triads, which are groups of three nodes, can also be useful for describing the structure of teams' social networks~\cite{zverevaTriadCensusUsage2016}. The triad census of a network is a set containing the frequencies of each triad type. Triads are valuable as they represent relationship structures. The presence of many triads with zero edges indicates that many members are not communicating with each other, while many triads with two edges indicates that members often communicate in pairs, and rarely communicate with members outside these pairs. This is considered particularly meaningful when analysing networks with few nodes~\cite{frankSurveyStatisticalMethods1981}, making it an appropriate characteristic for analysing teams of software developers. Finding the triad census of a network also contributes to calculating other network characteristics, like transitivity~\cite{newmanClusteringPreferentialAttachment2001}, so this is a useful first step. 

A more sophisticated analysis technique, socio-technical congruence, gives insight into how well a team's communication meets its coordination needs, which are identified through file changes in the project's repository~\cite{cataldoSociotechnicalCongruenceFramework2008}. If two developers change the same file, or files that are dependent on each other, then there is a need for coordination between those developers. The result of applying socio-technical congruence to a project is a score for each member representing the proportion of teammates they communicated with, of those they had a coordination need with. 

In professional software teams, congruence scores were positively correlated with productivity on both a team and individual level~\cite{cataldoIdentificationCoordinationRequirements2006}. Further, more productive team members improved their congruence scores over time. Conversely, MacKellar's examination of students' congruence scores in a team project showed that students' congruence scores did not increase over time~\cite{b.k.mackellarAnalyzingCoordinationStudents2013}. While the sample size for this study was small, the results suggest that congruence score trends may be less reflective of students' productivity than industry developers. Sierra et al. suggest this is a result of students being inexperienced with version control systems~\cite{sierraSystematicMappingStudy2018}. However, the student with the highest congruence score worked on a central piece of the project and mentored other students, indicating that we may identify team members who act as communication focal points through this type of analysis.


 We intend to first evaluate SNA techniques for identifying communication areas of improvement, contributing to our understanding of the suitability of these in the context of student software engineering teams. We then intend to contribute features to the project management tool that guide students towards improving their communication and coordination.


\section{Proposed Method}
\subsection{Context}
We will conduct our research in the context of SENG302, a 300-level two-semester software engineering group project at the University of Canterbury. Students work in teams of 6-8 members to develop a software application following the Scrum framework. SENG302 is divided into Sprints, which are each 2-3 term weeks long. Students attend formal sessions, including the Sprint Planning, Daily Scrums, Sprint Review, demonstration, and Sprint Retrospective sessions. Students submit weekly self-reflections, and each Sprint they submit feedback for their teammates. These discuss what each person did well, what they could improve, and the actions they can take to do so. Peer-feedback also includes ratings on a Likert scale from one to five for each teammate's contributions to the project through skills like testing and communication.

\subsection{Study 1}
Our first study focuses on answering RQ1 (\RQone). We applied SNA techniques identified from a literature survey to messaging and task allocation data from 2023 SENG302 teams, and compared the results with teams' project success (stories and story points passed), and intra-team communication scores. From this, triad census emerged as a promising indicator of areas of improvement in teams' communication~\cite{clarke2025improvingsoftwareengineeringteam}.

\subsection{Studies 2 and 3}
To evaluate RQ2 (\RQtwo) and RQ3 (\RQthree), we need to identify and evaluate features we can integrate into project management tools to influence teams' communication. We will first conduct a literature review, which will inform the implementation of features in the SENG302 project management tool. These features will be piloted on ex-SENG302 students, updated, and provided to the 2025 SENG302 cohort for the second semester. Based on the effects of the features on teams' communication and productivity, we will fine-tune the implementation, and provide a new version, or multiple versions, to the subsequent SENG302 cohort.

\subsection{Next Steps}
Our next steps are to evaluate more SNA techniques, and combine communication data sources for a more robust representation of team communication. So far, we have used work logs, Git commits, and written communication separately, but combining these may be more accurate. We will also confirm that triad census and any other promising techniques produce consistent results with data from the SENG302 2023 and 2024 cohorts before conducting a literature review on interventions.

\clearpage
\balance
\bibliographystyle{IEEEtran}
\bibliography{IEEEabrv,refs.bib}

\begin{thebibliography}{10}
\providecommand{\url}[1]{#1}
\csname url@samestyle\endcsname
\providecommand{\newblock}{\relax}
\providecommand{\bibinfo}[2]{#2}
\providecommand{\BIBentrySTDinterwordspacing}{\spaceskip=0pt\relax}
\providecommand{\BIBentryALTinterwordstretchfactor}{4}
\providecommand{\BIBentryALTinterwordspacing}{\spaceskip=\fontdimen2\font plus
\BIBentryALTinterwordstretchfactor\fontdimen3\font minus \fontdimen4\font\relax}
\providecommand{\BIBforeignlanguage}[2]{{%
\expandafter\ifx\csname l@#1\endcsname\relax
\typeout{** WARNING: IEEEtran.bst: No hyphenation pattern has been}%
\typeout{** loaded for the language `#1'. Using the pattern for}%
\typeout{** the default language instead.}%
\else
\language=\csname l@#1\endcsname
\fi
#2}}
\providecommand{\BIBdecl}{\relax}
\BIBdecl

\bibitem{curtisFieldStudySoftware1988}
\BIBentryALTinterwordspacing
B.~Curtis, H.~Krasner, and N.~Iscoe, ``A field study of the software design process for large systems,'' \emph{Communications of the ACM}, vol.~31, no.~11, pp. 1268--1287. [Online]. Available: \url{https://dl.acm.org/doi/10.1145/50087.50089}
\BIBentrySTDinterwordspacing

\bibitem{hoeglTeamworkQualitySuccess2001}
\BIBentryALTinterwordspacing
M.~Hoegl and H.~G. Gemuenden, ``Teamwork {{Quality}} and the {{Success}} of {{Innovative Projects}}: {{A Theoretical Concept}} and {{Empirical Evidence}},'' \emph{Organization Science}, vol.~12, no.~4, pp. 435--449. [Online]. Available: \url{https://pubsonline.informs.org/doi/10.1287/orsc.12.4.435.10635}
\BIBentrySTDinterwordspacing

\bibitem{mensahImpactTeamworkQuality2024}
A.~Mensah, ``The {{Impact}} of {{Teamwork Quality}} ({{TWQ}}) on {{Agile Software Development Team Performance}},'' \emph{International Journal of Project Management}, vol.~6, pp. 25--51.

\bibitem{krautCoordinationSoftwareDevelopment1995}
\BIBentryALTinterwordspacing
R.~E. Kraut and L.~A. Streeter, ``Coordination in software development,'' \emph{Communications of the ACM}, vol.~38, no.~3, pp. 69--81. [Online]. Available: \url{https://dl.acm.org/doi/10.1145/203330.203345}
\BIBentrySTDinterwordspacing

\bibitem{r.noelExploringCollaborativeWriting2018}
{R. Noel}, {F. Riquelme}, {R. M. Lean}, {E. Merino}, {C. Cechinel}, {T. S. Barcelos}, {R. Villarroel}, and {R. Munoz}, ``Exploring {{Collaborative Writing}} of {{User Stories With Multimodal Learning Analytics}}: {{A Case Study}} on a {{Software Engineering Course}},'' \emph{IEEE Access}, vol.~6, pp. 67\,783--67\,798.

\bibitem{strodeTeamworkEffectivenessModel2022a}
\BIBentryALTinterwordspacing
D.~Strode, T.~Dingsøyr, and Y.~Lindsjorn, ``A teamwork effectiveness model for agile software development,'' \emph{Empirical Software Engineering}, vol.~27, no.~2, p.~56. [Online]. Available: \url{https://link.springer.com/10.1007/s10664-021-10115-0}
\BIBentrySTDinterwordspacing

\bibitem{salasHowCanYou1997}
E.~Salas, J.~A. Cannon-Bowers, and J.~H. Johnston, \emph{How Can You Turn a Team of Experts into an Expert Team? {{Emerging}} Training Strategies. {{I CE Zsambok}} \& {{G}}. {{Klein}} (Red.), {{Naturalistic Decision Making}} (s. 359–370)}.\hskip 1em plus 0.5em minus 0.4em\relax Mahwah, NJ, USA: Erlbaum Associates.

\bibitem{omalleyAnalysisSocialNetworks2008}
\BIBentryALTinterwordspacing
A.~J. O’Malley and P.~V. Marsden, ``The analysis of social networks,'' \emph{Health Services and Outcomes Research Methodology}, vol.~8, no.~4, pp. 222--269. [Online]. Available: \url{http://link.springer.com/10.1007/s10742-008-0041-z}
\BIBentrySTDinterwordspacing

\bibitem{sparroweSocialNetworksPerformance2001}
\BIBentryALTinterwordspacing
R.~T. Sparrowe, R.~C. Liden, S.~J. Wayne, and M.~L. Kraimer, ``Social {{Networks}} and the {{Performance}} of {{Individuals}} and {{Groups}},'' \emph{The Academy of Management Journal}, vol.~44, no.~2, pp. 316--325. [Online]. Available: \url{http://www.jstor.org/stable/3069458}
\BIBentrySTDinterwordspacing

\bibitem{seersTeammemberExchangeQuality1989}
\BIBentryALTinterwordspacing
A.~Seers, ``Team-member exchange quality: {{A}} new construct for role-making research,'' \emph{Organizational Behavior and Human Decision Processes}, vol.~43, no.~1, pp. 118--135. [Online]. Available: \url{https://www.sciencedirect.com/science/article/pii/0749597889900605}
\BIBentrySTDinterwordspacing

\bibitem{otteSocialNetworkAnalysis2002}
E.~Otte and R.~Rousseau, ``Social {{Network Analysis}}: {{A Powerful Strategy}}, also for the {{Information Sciences}},'' \emph{Journal of Information Science}, vol.~28, pp. 441--453.

\bibitem{zhangUnderstandingStudentTeachers2022}
\BIBentryALTinterwordspacing
S.~Zhang, Q.~Gao, M.~Sun, Z.~Cai, H.~Li, Y.~Tang, and Q.~Liu, ``Understanding student teachers’ collaborative problem solving: {{Insights}} from an epistemic network analysis ({{ENA}}),'' \emph{Computers \& Education}, vol. 183, p. 104485. [Online]. Available: \url{https://linkinghub.elsevier.com/retrieve/pii/S0360131522000562}
\BIBentrySTDinterwordspacing

\bibitem{mathieuTeamEffectiveness199720072008}
\BIBentryALTinterwordspacing
J.~Mathieu, M.~T. Maynard, T.~Rapp, and L.~Gilson, ``Team {{Effectiveness}} 1997-2007: {{A Review}} of {{Recent Advancements}} and a {{Glimpse Into}} the {{Future}},'' \emph{Journal of Management}, vol.~34, no.~3, pp. 410--476. [Online]. Available: \url{http://journals.sagepub.com/doi/10.1177/0149206308316061}
\BIBentrySTDinterwordspacing

\bibitem{hughesRemotePairProgramming2020}
\BIBentryALTinterwordspacing
J.~Hughes, A.~Walshe, B.~Law, and B.~Murphy, ``Remote {{Pair Programming}}:,'' in \emph{Proceedings of the 12th {{International Conference}} on {{Computer Supported Education}}}.\hskip 1em plus 0.5em minus 0.4em\relax {SCITEPRESS - Science and Technology Publications}, pp. 476--483. [Online]. Available: \url{http://www.scitepress.org/DigitalLibrary/Link.aspx?doi=10.5220/0009582904760483}
\BIBentrySTDinterwordspacing

\bibitem{mcdowellPairProgrammingImproves2006}
\BIBentryALTinterwordspacing
C.~McDowell, L.~Werner, H.~E. Bullock, and J.~Fernald, ``Pair programming improves student retention, confidence, and program quality,'' \emph{Communications of the ACM}, vol.~49, no.~8, pp. 90--95. [Online]. Available: \url{https://dl.acm.org/doi/10.1145/1145287.1145293}
\BIBentrySTDinterwordspacing

\bibitem{zverevaTriadCensusUsage2016}
\BIBentryALTinterwordspacing
O.~M. Zvereva, ``Triad census usage for communication network analysis,'' in \emph{{{CEUR Workshop Proceedings}}}, vol. 1710.\hskip 1em plus 0.5em minus 0.4em\relax CEUR-WS, pp. 378--389. [Online]. Available: \url{https://ceur-ws.org/Vol-1710/paper38.pdf}
\BIBentrySTDinterwordspacing

\bibitem{frankSurveyStatisticalMethods1981}
\BIBentryALTinterwordspacing
O.~Frank, ``A {{Survey}} of {{Statistical Methods}} for {{Graph Analysis}},'' \emph{Sociological Methodology}, vol.~12, p. 110. [Online]. Available: \url{https://www.jstor.org/stable/270740?origin=crossref}
\BIBentrySTDinterwordspacing

\bibitem{newmanClusteringPreferentialAttachment2001}
\BIBentryALTinterwordspacing
M.~E.~J. Newman, ``Clustering and preferential attachment in growing networks,'' \emph{Physical Review E}, vol.~64, no.~2, p. 025102. [Online]. Available: \url{https://link.aps.org/doi/10.1103/PhysRevE.64.025102}
\BIBentrySTDinterwordspacing

\bibitem{cataldoSociotechnicalCongruenceFramework2008}
\BIBentryALTinterwordspacing
M.~Cataldo, J.~D. Herbsleb, and K.~M. Carley, ``Socio-technical congruence: A framework for assessing the impact of technical and work dependencies on software development productivity,'' in \emph{Proceedings of the {{Second ACM-IEEE}} International Symposium on {{Empirical}} Software Engineering and Measurement}.\hskip 1em plus 0.5em minus 0.4em\relax ACM, pp. 2--11. [Online]. Available: \url{https://dl.acm.org/doi/10.1145/1414004.1414008}
\BIBentrySTDinterwordspacing

\bibitem{cataldoIdentificationCoordinationRequirements2006}
\BIBentryALTinterwordspacing
M.~Cataldo, P.~A. Wagstrom, J.~D. Herbsleb, and K.~M. Carley, ``Identification of coordination requirements: Implications for the {{Design}} of collaboration and awareness tools,'' in \emph{Proceedings of the 2006 20th Anniversary Conference on {{Computer}} Supported Cooperative Work}.\hskip 1em plus 0.5em minus 0.4em\relax ACM, pp. 353--362. [Online]. Available: \url{https://dl.acm.org/doi/10.1145/1180875.1180929}
\BIBentrySTDinterwordspacing

\bibitem{b.k.mackellarAnalyzingCoordinationStudents2013}
{B. K. MacKellar}, ``Analyzing coordination among students in a software engineering project course,'' in \emph{2013 26th {{International Conference}} on {{Software Engineering Education}} and {{Training}} ({{CSEE}}\&{{T}})}, pp. 279--283.

\bibitem{sierraSystematicMappingStudy2018}
\BIBentryALTinterwordspacing
J.~M. Sierra, A.~Vizcaíno, M.~Genero, and M.~Piattini, ``A systematic mapping study about socio-technical congruence,'' \emph{Information and Software Technology}, vol.~94, pp. 111--129. [Online]. Available: \url{https://www.sciencedirect.com/science/article/pii/S0950584916302798}
\BIBentrySTDinterwordspacing

\bibitem{clarke2025improvingsoftwareengineeringteam}
A.~Clarke, T.~Mitrović, and F.~Gilson, ``Improving software engineering team communication through stronger social networks,'' 2025, preprint arxiv:2502.01923.

\end{thebibliography}

\end{document}